\documentclass[10pt,aps,prl,twocolumn,fleqn,superscriptaddress,nofootinbib,preprintnumbers,nobalancelastpage]{revtex4-1}
\usepackage{amsmath,amssymb,pstricks,graphicx,xspace,subfigure,ulem}
 
\def\gtap{\raisebox{-.6ex}{\rlap{$\,\sim\,$}} \raisebox{.4ex}{$\,>\,$}} 
 
\newcommand\as{\alpha_{\mathrm{S}}} 
\def\ptveto{p^{\mathrm{veto}}_{T,\mathrm{bjet}}}
\def\ptb{p_{T,\mathrm{bjet}}}

\begin{document}

\preprint{ZU-TH 29/14}
\preprint{MITP/14-053}

\title{$W^+W^-$ production at hadron colliders in NNLO QCD}

\author{T.~Gehrmann}
\affiliation{Physik-Institut, Universit{\"a}t Z{\"u}rich, CH-8057 Z{\"u}rich, Switzerland}

\author{M.~Grazzini\footnote{On leave of absence from INFN, 
Sezione di Firenze, Sesto Fiorentino, Florence, Italy.}}
\affiliation{Physik-Institut, Universit{\"a}t Z{\"u}rich, CH-8057 Z{\"u}rich, Switzerland}

\author{S.~Kallweit}
\affiliation{Physik-Institut, Universit{\"a}t Z{\"u}rich, CH-8057 Z{\"u}rich, Switzerland}

\author{P.~Maierh{\"o}fer}
\affiliation{Physik-Institut, Universit{\"a}t Z{\"u}rich, CH-8057 Z{\"u}rich, Switzerland}

\author{A.~von~Manteuffel}
\affiliation{PRISMA Cluster of Excellence, Institute of Physics,
Johannes Gutenberg University, D-55099 Mainz, Germany}

\author{S.~Pozzorini}
\affiliation{Physik-Institut, Universit{\"a}t Z{\"u}rich, CH-8057 Z{\"u}rich, Switzerland}

\author{D.~Rathlev}
\affiliation{Physik-Institut, Universit{\"a}t Z{\"u}rich, CH-8057 Z{\"u}rich, Switzerland}

\author{L.~Tancredi}
\affiliation{Physik-Institut, Universit{\"a}t Z{\"u}rich, CH-8057 Z{\"u}rich, Switzerland}

\begin{abstract}
Charged gauge boson pair production at the Large Hadron Collider allows
detailed probes of the 
fundamental structure of electroweak interactions.
We present precise theoretical predictions for on-shell $W^+W^-$
production that include, for the first time, QCD effects up to 
next-to-next-to-leading order in perturbation theory.
As compared to next-to-leading order,
the inclusive $W^+W^-$ cross section is enhanced by 9\% at $7$~TeV and 12\% at
$14$~TeV.
The residual perturbative uncertainty is at the 3\% level.  
The severe contamination of the $W^+W^-$ cross section due to top-quark
resonances is discussed in detail.  Comparing different definitions of
top-free $W^+W^-$ production in the four and five flavour number schemes, we
demonstrate that top-quark resonances can be separated from the inclusive
$W^+W^-$ cross section without significant loss of theoretical precision.

\end{abstract}

\maketitle

Vector boson pair production is among the most important electroweak 
processes at hadron colliders. It allows detailed studies of the 
gauge symmetry structure of electroweak interactions
and of the mechanism of electroweak symmetry breaking. Any deviation 
from Standard Model expectations in measured production rates and kinematical 
distributions of vector boson pairs or their decay products could provide 
first evidence for new-physics effects at the high-energy frontier. 
Vector boson pair production is moreover an important 
background in measurements of Higgs boson 
production~\cite{Aad:2012tfa,Chatrchyan:2012ufa} 
 and in direct searches for new particles.

Among the massive vector boson pair production reactions, $W^+W^-$ takes a special role, in 
having a larger cross section than $W^\pm Z$ and $ZZ$ production, 
while at the same time producing the 
most challenging
final state with $W^+W^-\to l^+\nu l^-\bar{\nu} $. Due to the presence of two neutrinos, 
it does not allow to reconstruct mass peaks, and
its control requires a 
very thorough understanding of the $W^+W^-$ signal and
its background contamination.
Various measurements of $W^+W^-$ hadroproduction have been carried out at the Tevatron and the LHC
(for some recent results see Refs.~\cite{Abazov:2011cb,CDFnote,ATLAS:2012mec,Chatrchyan:2013yaa,Chatrchyan:2013oev,ATLAS-CONF-2014-033}). 
The observation of a total 
$W^+W^-$ cross section at 8~TeV in excess of theoretical expectations has triggered 
intensive discussion~\cite{Curtin:2014zua,Kim:2014eva,Luo:2014hrk} 
about possible new-physics 
effects showing up here for the first time. In order to establish or refute this excess, it is mandatory 
to have a solid theoretical prediction (with a reliable estimate of its residual uncertainty) 
for $W^+W^-$ production. In this Letter, we bring this prediction to a new level of 
accuracy with the first-ever computation of next-to-next-to-leading order~(NNLO) QCD 
corrections to the inclusive $W^+W^-$ hadroproduction cross section.

Following the leading-order~(LO) estimate of the $W^+W^-$ cross section~\cite{Brown:1978mq},
next-to-leading order~(NLO) QCD corrections~\cite{Ohnemus:1991kk,Frixione:1993yp} 
were first evaluated by considering stable $W$ bosons.
The computation of the relevant one-loop helicity amplitudes~\cite{Dixon:1998py} 
allowed
complete NLO calculations~\cite{Campbell:1999ah,Dixon:1999di}, 
including
spin correlations and off-shell effects.
The loop-induced gluon fusion contribution, which is formally NNLO,
has been computed in Refs.~\cite{Dicus:1987dj,Glover:1988fe}. 
The corresponding leptonic decays have been
included in Refs.~\cite{Binoth:2005ua,Binoth:2006mf}, 
and, more recently,
the interference with the $gg\to H$ signal has been taken into account~\cite{Campbell:2011cu}.
Since the gluon-induced contribution is enhanced by the gluon luminosity, it is often assumed to
provide the bulk of the NNLO corrections.
NLO predictions for $W^+W^-$ production including the gluon-induced contribution,
the leptonic decay with spin correlations and off-shell effects have
been presented in Ref.~\cite{Campbell:2011bn}. 
The NLO QCD corrections to $W^+W^-+{\rm jet}$ production have been discussed in 
Refs.~\cite{Dittmaier:2007th,Campbell:2007ev,Dittmaier:2009un}, 
and even 
NLO results for 
$W^+W^-+2~{\rm jets}$ are available~\cite{Melia:2011dw,Greiner:2012im}. 
The effects of transverse-momentum \cite{Grazzini:2005vw,Wang:2013qua,Meade:2014fca}, jet veto \cite{Jaiswal:2014yba} and threshold \cite{Dawson:2013lya} resummation
for $W^+W^-$ production have also been investigated.
The electroweak (EW) corrections to this process have been computed in 
Refs.~\cite{Bierweiler:2012kw,Baglio:2013toa,Billoni:2013aba}.
Detailed Monte Carlo simulations of $e^+\nu_e\mu^-\bar\nu_\mu$ production in association with up to 
one jet at NLO have been presented in Ref.~\cite{Cascioli:2013gfa}.

In this Letter we report on the first calculation of the inclusive
production of on-shell $W$-boson pairs at hadron colliders in NNLO QCD.  
The calculation parallels the one presented for $Z$-boson pairs in
Ref.~\cite{Cascioli:2014yka}, 
but differs from it on one important aspect.
The higher-order QCD corrections to $W^+W^-$ production include partonic
channels with $b$-quarks in the final
state, which lead to a subtle interplay between $W^+W^-$ and top 
production processes~\cite{Dittmaier:2007th,Cascioli:2013gfa}.
In the five flavour number scheme~(FNS), where $b$-quarks are included in the parton
distribution functions and their mass is set to zero, the presence of real
$b$-quark emission is crucial in order to cancel collinear singularities that
arise from $g\to b\bar b$ splittings in the virtual corrections.  At the same time,
the occurrence of $Wb$ pairs in the real-emission matrix elements induces
top-quark resonances that lead to a problematic contamination of $W^+W^-$
production.  The problem starts with the NLO cross section, which receives a
contribution of about 30~(60)\% at 7~(14)~TeV from $pp\to W^\pm t\to W^+W^-b$, and at NNLO
the appearance of doubly resonant $pp\to t{\bar t}\to W^+W^-b\bar b$ channels
enhances the $W^+W^-$ cross section by about a factor 4~(8).

This huge contamination calls for a theoretical definition of $W^+W^-$
production where top contributions are completely subtracted, similarly as
in the experimental measurements of the $W^+W^-$ cross
section~\cite{Abazov:2011cb,CDFnote,ATLAS:2012mec,Chatrchyan:2013yaa,Chatrchyan:2013oev,ATLAS-CONF-2014-033}. 
However, the need of cancelling collinear $g\to b\bar b$ singularities does
not allow for a trivial separation of $W^+W^-$ and top production in the 5FNS. 
To address this issue two different definitions of $W^+W^-$ production will be
adopted and compared in this Letter.  The first definition is based on the
4FNS.  In this case, since  $b$-quarks are massive and collinear
divergences are not present, we define top-free $W^+W^-$ production 
by simply omitting $b$-quark emissions.
%
%
Alternatively, we will adopt a
5FNS definition of $W^+W^-$ production, where $b$-quark emissions are included. 
In this case, for a consistent separation of the $tW$ and
$t\bar t$ contributions we will introduce 
a top subtraction based on the scaling behaviour of the (N)NLO cross section 
in the limit of vanishing top-quark width.
The comparison of 4FNS and 5FNS predictions
will permit us to quantify the theoretical ambiguities 
inherent in a top-free definition of the $W^+W^-$ cross section at NNLO.


The computation of NNLO corrections 
requires the evaluation of the tree-level
scattering amplitudes with two additional (unresolved) partons, of the one-loop amplitudes with 
one additional parton, and of the one-loop-squared and
two-loop corrections to the Born subprocess $q{\bar q}\to W^+W^-$.
In our calculation, all required tree and one-loop matrix elements are automatically 
generated with {\sc OpenLoops}~\cite{Cascioli:2011va}, 
which implements a fast numerical recursion 
for the calculation of NLO scattering amplitudes within the Standard Model. 
For the numerically stable evaluation of tensor integrals
we rely on the {\sc Collier} library~\cite{collier}, 
which is based on the Denner--Dittmaier reduction techniques~\cite{Denner:2002ii,Denner:2005nn} 
and the scalar integrals of~\cite{Denner:2010tr}.
To check and further improve the numerical stability of exceptional phase space points 
the quadruple precision implementation of the OPP method~\cite{Ossola:2006us} 
in {\sc CutTools}~\cite{Ossola:2007ax} 
is employed in combination with {\sc OneLOop}~\cite{vanHameren:2010cp}.
Following the recent computation of the relevant two-loop master 
integrals~\cite{Gehrmann:2013cxs,Henn:2014lfa,Gehrmann:2014bfa,Caola:2014lpa,Anastasiou:2014}
the last missing contribution,
the genuine two-loop correction to the $W^+W^-$ amplitude,
has been computed by some of us 
and will be reported elsewhere~\cite{inprep},
thereby improving upon earlier results in the high-energy 
limit~\cite{Chachamis:2008yb}.
In the two-loop correction, contributions involving a top-quark loop
are neglected. For the numerical evaluation of the multiple
polylogarithms in the two-loop expressions we employ the
implementation~\cite{Vollinga:2004sn} 
in the {\sc GiNaC}~\cite{Bauer:2000cp}
library.

The implementation of the various scattering amplitudes in a complete NNLO calculation is
a non-trivial task due to the presence of infrared~(IR) singularities at
intermediate stages of the calculation that prevent a straightforward application of numerical techniques.
To handle and cancel these singularities at NNLO
we employ the $q_T$ subtraction method~\cite{Catani:2007vq}.
This approach determines the IR singular 
behaviour of real radiation contributions from the resummation of 
logarithmically-enhanced
contributions to $q_T$ distributions. In the case of the production of a colourless high-mass system,
the $q_T$ subtraction method is fully developed \cite{Bozzi:2005wk,Catani:2013tia}, thanks to the computation of the relevant hard-collinear coefficients
\cite{Catani:2011kr, Catani:2012qa}, later confirmed 
with an independent calculation in the framework of 
Soft-Collinear Effective Theory~(SCET) \cite{Gehrmann:2012ze,Gehrmann:2014yya}.
The $q_T$ subtraction method has been used for the computation of NNLO corrections to
several hadronic processes \cite{Catani:2007vq,Catani:2009sm,Ferrera:2011bk,Catani:2011qz,Grazzini:2013bna,Cascioli:2014yka,Ferrera:2014lca}.

We have performed our NNLO calculation for $W^+W^-$ production
starting from a computation of the $d{\sigma}^{W^+W^-+{\rm jet}}_{NLO}$ cross
section with the dipole-subtraction method~\cite{Catani:1996jh,Catani:1996vz}.  
The numerical calculation employs the
generic Monte Carlo program that was developed for
Refs.~\cite{Grazzini:2013bna,Cascioli:2014yka}.  
Although the $q_T$
subtraction method and our implementation are suitable to perform a fully
exclusive computation of $W^+W^-$ production including the leptonic decays and
the corresponding spin correlations, in this Letter we restrict ourselves to
the inclusive production of on-shell $W$ bosons.

In the following we present LO, NLO and NNLO predictions for $pp\to W^+W^-+X$ 
with $\sqrt{s}$ ranging from 7 to 14~TeV.  
We use the MSTW2008 
sets of parton distributions
with four~\cite{Martin:2010db} or five~\cite{Martin:2009iq} active flavours.
Parton densities and $\as$ are evaluated at each corresponding order, i.e.~we use
$(n+1)$-loop $\as$ at N$^n$LO, with $n=0,1,2$.
The default renormalization ($\mu_R$) and factorization ($\mu_F$) scales are
set to $\mu_R=\mu_F=m_W$, and to assess scale uncertainties they are varied in the range
$0.5\,m_W<\mu_{R,F} <2\,m_W$ with $0.5<\mu_F/\mu_R<2$.
In the 4FNS we use $m_b=4.75$ GeV, while in the 5FNS $b$-quarks are massless. 
The electroweak parameters are defined in the
$G_\mu$ scheme, 
with $G_F = 1.16639\times 10^{-5}$~GeV$^{-2}$, $m_W=80.399$ GeV, and $m_Z =
91.1876$~GeV.  
Our NLO and NNLO predictions involve resonant top quarks 
and off-shell Higgs bosons, and for the respective mass and width 
parameters we use $m_t=173.2$ GeV, $\Gamma_t=1.443$ GeV, 
$m_H=125$ GeV and $\Gamma_H=4.09$ MeV.
Higgs contributions are included via squared one-loop amplitudes in the 
$gg\to H^* \to W^+W^-$ channel, but are strongly suppressed by the off-shellness of the 
Higgs boson.

\renewcommand{\baselinestretch}{1.6}
\begin{table}[ht]
\begin{center}
\begin{tabular}{|c| c| c| c| c|}
\hline
$\frac{\sqrt{s}}{\mathrm{TeV}}$
& $\sigma_{LO}$ 
& $\sigma_{NLO}$ 
& $\sigma_{NNLO}$ 
& $\sigma_{gg\to H\to WW^*}$ 
\\ [0.5ex]
\hline
7 & $29.52^{+1.6\%}_{-2.5\%}$ & $45.16^{+3.7\%}_{-2.9\%}$ & $49.04^{+2.1\%}_{-1.8\%}$ & $3.25^{+7.1\%}_{-7.8\%}$\\
\hline
8 & $35.50^{+2.4\%}_{-3.5\%}$ & $54.77^{+3.7\%}_{-2.9\%}$ & $59.84^{+2.2\%}_{-1.9\%}$ & $4.14^{+7.2\%}_{-7.8\%}$\\
\hline
13 & $67.16^{+5.5\%}_{-6.7\%}$ & $106.0^{+4.1\%}_{-3.2\%}$ & $118.7^{+2.5\%}_{-2.2\%}$ & $9.44^{+7.4\%}_{-7.9\%}$\\
\hline
14 & $73.74^{+5.9\%}_{-7.2\%}$ & $116.7^{+4.1\%}_{-3.3\%}$ & $131.3^{+2.6\%}_{-2.2\%}$ & $10.64^{+7.5\%}_{-8.0\%}$\\
\hline
\end{tabular}
\end{center}

\renewcommand{\baselinestretch}{1.0}
\caption{LO, NLO and NNLO cross sections (in picobarn) for 
on-shell $W^+W^-$ production in the 4FNS
and reference results for $gg\to H \to WW^*$~ from Ref.~\cite{Heinemeyer:2013tqa}.
} 
\label{table4F}
\end{table}


In Table~\ref{table4F} we present LO, NLO and NNLO predictions for inclusive
$W^+W^-$ production in the 4FNS, where top contributions are
removed by omitting $b$-quark emissions.
We see that at 7~(14)~TeV
the LO predictions receive a positive NLO shift of 53~(58)\%, and the NNLO
corrections induce a further enhancement of 9~(12)\%.
The decent perturbative convergence is contrasted by the observation that 
the scale uncertainty does not significantly decrease when moving from
LO to NLO and NNLO.  Moreover, the NNLO (NLO) corrections turn out to exceed the scale
uncertainty of the NLO (LO) predictions by up to a factor 3~(34).  
The fact that LO and NLO scale variations underestimate higher-order effects
can be attributed to the fact that the gluon--quark and gluon--gluon induced
partonic channels, which yield a sizable contribution to the $W^+W^-$ cross
section, appear only beyond LO and NLO, respectively.  The NNLO is the first order
at which all partonic channels are contributing.  The NNLO scale dependence, which
amounts to about 3\%, can thus be considered a realistic estimate of the
theoretical uncertainty due to missing higher-order effects.

\begin{figure}[ht]
\begin{center}
\hspace{-5mm}\includegraphics[width=0.43\textwidth,angle=0]{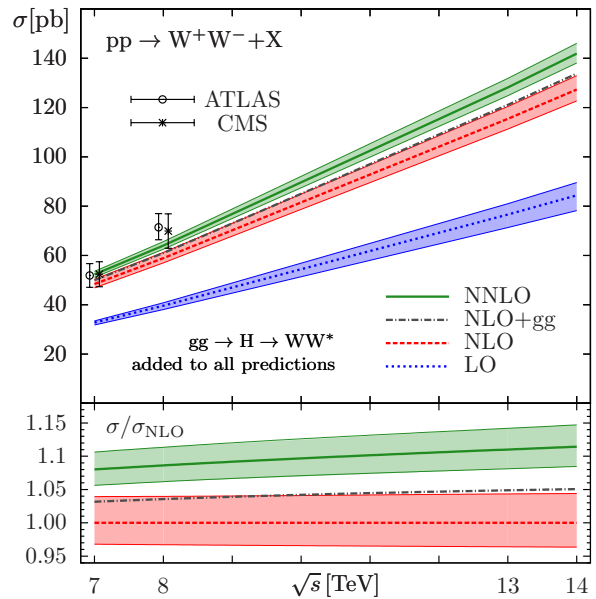}
\end{center}
\vspace{-4mm}
\caption{The on-shell $W^+W^-$ cross section in the 4FNS at 
LO (dots), NLO (dashes), NLO+gg (dot dashes) and NNLO (solid) 
combined with $gg\to H \to WW^*$
is compared to recent ATLAS and CMS measurements~\cite{ATLAS:2012mec,Chatrchyan:2013yaa,Chatrchyan:2013oev,ATLAS-CONF-2014-033}. 
In the lower panel NNLO and NLO+gg results are normalized to NLO predictions.
The bands describe scale variations.}
\label{figure_sqrts}
\end{figure}

In Figure~\ref{figure_sqrts}, theoretical predictions in the 4FNS are compared to 
CMS and ATLAS measurements at 7 and
8~TeV~\cite{ATLAS:2012mec,Chatrchyan:2013yaa,Chatrchyan:2013oev,ATLAS-CONF-2014-033}.
For a consistent comparison, 
our results for on-shell
$W^+W^-$ production are combined with the $gg\to H \to WW^*$ cross
sections reported in Table \ref{table4F}.  It turns out
that the inclusion of the NNLO corrections leads to an excellent description
of the data at 7~TeV and decreases the significance of the observed excess
at 8~TeV.  
In the lower frame of Figure~\ref{figure_sqrts}, predictions and scale
variations at NNLO are compared to NLO ones, and also the individual
contribution of the $gg\to W^+W^-$ channel is shown. Using NNLO parton distributions throughout, the loop induced
gluon fusion contribution is only about 35\% of the total NNLO correction.
\begin{figure*}[ht]
\begin{center}
\hspace{-4mm}
\includegraphics[width=0.43\textwidth,angle=0]{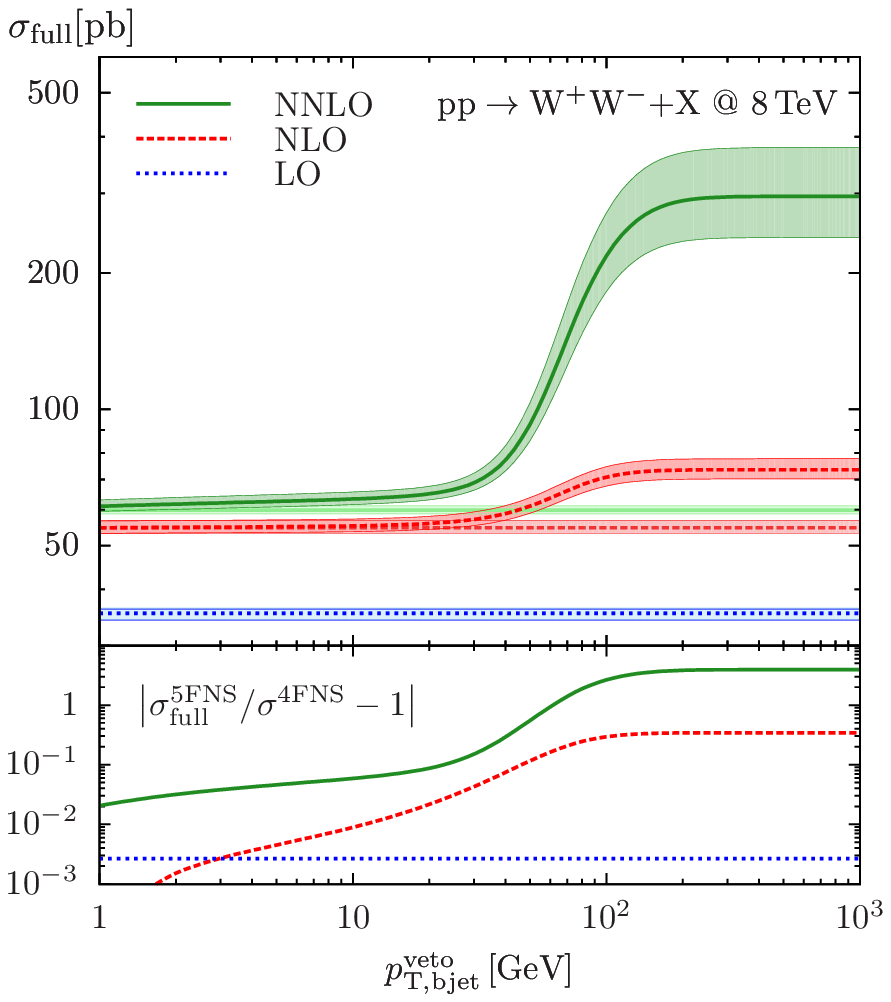}
\hspace{8mm}
\includegraphics[width=0.43\textwidth,angle=0]{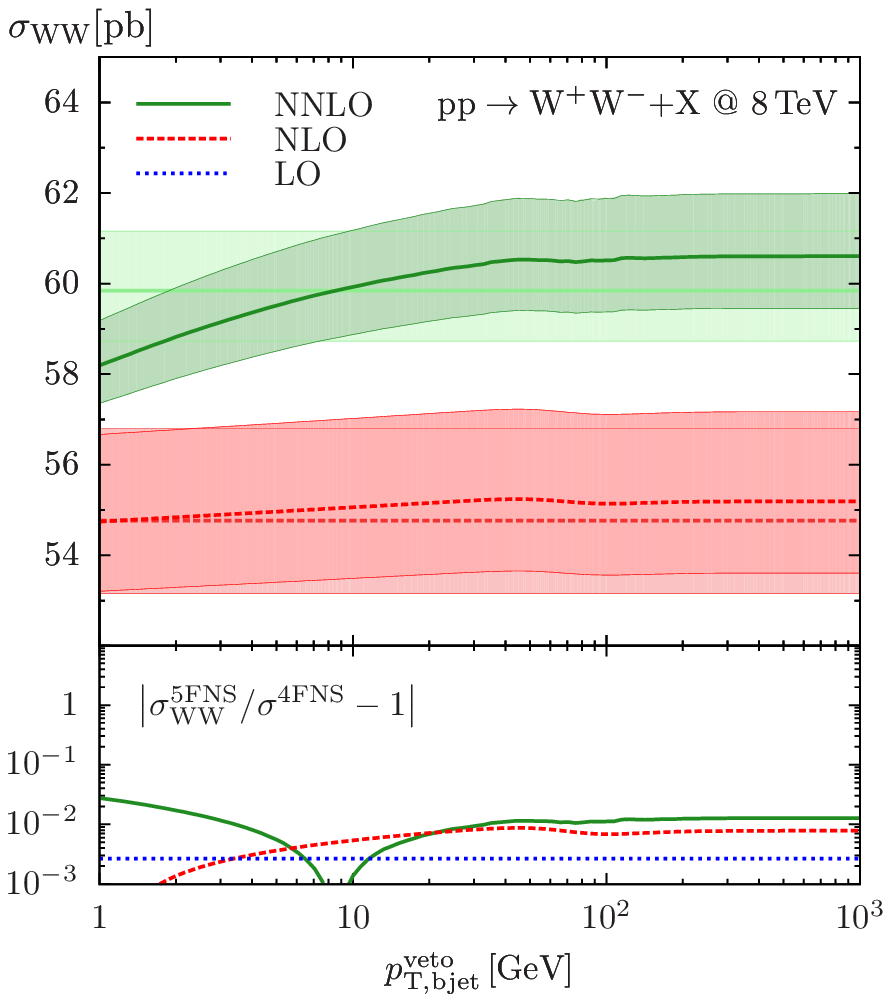}
\end{center}
\vspace{-4mm}
\caption{
The $pp\to W^+W^-$ cross section in the 5FNS at $\sqrt{s}=8$ TeV  
is plotted versus a $b$-jet veto, $\ptb <\ptveto$,
and compared to results in the 4FNS (which are $\ptveto$ independent).
Full 5FNS results (left plot) are contrasted with top-subtracted 
5FNS predictions (right plot).
The relative agreement between 5FNS and 4FNS results is displayed in the
lower frames. 
Jets are defined using the anti-$k_T$
algorithm~\cite{Cacciari:2008gp} with $R=0.4$, and in order to guarantee
the cancellations of final-state collinear singularities, $b\bar b$ pairs
that are recombined by the jet algorithm are not vetoed.
}
\label{figure_5f}
\end{figure*}

In the light of the small scale dependence of the 4FNS NNLO cross section,
the ambiguities associated with the definition of a
top-free $W^+W^-$ cross section and its sensitivity to the choice of the
FNS might represent a significant source of theoretical
uncertainty at NNLO. In particular, the omission of
$b$-quark emissions in our 4FNS definition of the $W^+W^-$ cross section implies
potentially large logarithms of $m_b$ in the transition from the 4FNS to the
5FNS.  To quantify this kind of uncertainties, we study the NNLO $W^+W^-$ cross 
section in the 5FNS and introduce a subtraction of its top contamination that allows for a
consistent comparison between the two FNSs.
An optimal definition of $W^+W^-$ production in the 5FNS requires maximal
suppression of the top resonances in the $pp\to W^+W^-b$ and $pp\to W^+W^-b\bar{b}$
channels. At the same time, the
cancellation of collinear singularities associated with massless $g\to
b\bar b$ splittings requires a sufficient level of inclusiveness.
The difficulty of fulfilling both requirements is
clearly illustrated in Figure~\ref{figure_5f} (left), where 5FNS 
predictions are plotted versus a $b$-jet veto that rejects $b$-jets with $\ptb>\ptveto$ over the whole rapidity range,
and are compared to 4FNS results.
In the inclusive limit, $\ptveto\to\infty$, the higher-order corrections in
the 5FNS suffer from a huge top contamination.  At 7~(14)~TeV the resulting
relative enhancement with respect to the 4FNS amounts to about 30~(60)\% at
NLO and a factor 4~(8) at NNLO.  In principle, it can be suppressed through the
$b$-jet veto.  However, for natural jet veto values around 30~GeV the top
contamination remains larger than 10\% of the $W^+W^-$ cross section, and a
complete suppression of the top contributions requires a veto of the order
of 1~GeV. Moreover, as $\ptveto\to 0$, the (N)NLO cross section does not 
approach a constant, but, starting from $\ptveto\sim 10$ GeV, it displays a logarithmic slope due to 
singularities associated
with initial state $g\to b\bar b$ splittings.  
This sensitivity to the
jet-veto parameters represents a theoretical ambiguity at the several percent level,
which is inherent in
the definition of top-free $W^+W^-$ production based on a $b$-jet veto.

\newcommand{\rd}{\mathrm{d}}
\newcommand{\sfull}{\sigma^{5F}_{\mathrm{full}}}
\newcommand{\stt}{\sigma^{5F}_{t\bar t}}
\newcommand{\stw}{\sigma^{5F}_{tW}}
\newcommand{\sww}{\sigma^{5F}_{WW}}
\newcommand{\xit}{\xi_t}
\newcommand{\Gt}{\Gamma_t}
\newcommand{\Gtp}{\Gamma^{\mathrm{phys}}_t}

To circumvent this problem we will adopt an alternative definition of the
$W^+W^-$ cross section in the 5FNS, where resonant top contributions are
subtracted along the lines of Refs.~\cite{Denner:2012yc,Cascioli:2013wga} by
exploiting their characteristic scaling behaviour in the limit of vanishing
top-quark width. The idea is that
doubly (singly) resonant contributions feature a
quadratic (linear) dependence on $1/\Gt$, while top-free $W^+W^-$ contributions
are not enhanced at small $\Gt$. 
Using this scaling property, the 
$t\bar t$, $tW^\pm$ and (top-free) $W^+W^-$ components in the 5FNS 
are  determined from high-statistics evaluations of the 5FNS 
cross section at different values of $\Gt$. 
The 5FNS top-free $W^+W^-$ cross section $\sww$, defined in this way, 
is presented in Figure~\ref{figure_5f} (right) for
$\sqrt{s}=8$~TeV.
Its dependence on the $b$-jet veto demonstrates the 
consistency of the employed top subtraction:
at $\ptveto\to 0$
we clearly observe the above-mentioned QCD singularity 
from initial-state $g\to b\bar b$, 
while for $\ptveto\gtap 10$~GeV, consistently with the absence of top contamination, 
$\sww$ is almost insensitive to the veto. 
Thus the inclusive limit of $\sww$ can be used as a precise theoretical definition of $W^+W^-$ 
production
in the 5FNS, and compared to the 4FNS.  The agreement between the two schemes turns out to be
at the level of 1~(2)\% at 7~(14)~TeV, and this finding puts our NNLO results 
and their estimated uncertainty on a firm theoretical ground.

In summary, we have presented the first NNLO calculation 
of the total $W^+W^-$ production cross section at the LHC.
The $W^+W^-$ signature is of crucial importance to precision tests of
the 
fundamental structure of
electroweak interactions and provides an important background in Higgs
boson studies and searches for new physics.  
Introducing consistent 
 theoretical definitions of $W^+W^-$ production in the
four and five flavour number schemes, we have demonstrated that the huge top
contamination of the $W^+W^-$ signal can be subtracted without significant
loss of theoretical precision.
The NNLO corrections to $W^+W^-$ production increase from 9\% at 7~TeV to
12\% at 14~TeV, with an estimated 3\% residual uncertainty from missing
contributions beyond NNLO.  Gluon fusion amounts to about 35\% of the total
NNLO contribution.
The inclusion of the newly computed NNLO corrections provides an excellent
description of recent measurements of the $W^+W^-$ cross section at 7~TeV and
diminishes the significance of an observed excess at 8~TeV.  In the near future
more differential studies at NNLO, including leptonic decays and
off-shell effects, will open the door to high-precision phenomenology with
$W^+W^-$ final states.
\begin{acknowledgments}
We would like to thank A.~Denner, S.~Dittmaier and L.~Hofer for providing us with the {\sc Collier}
library. S.~K., P.~M. and S.~P. would like to thank the Munich Institute for
Astro- and Particle Physics (MIAPP) of the DFG cluster of excellence
"Origin and Structure of the Universe" for the hospitality during 
the completion of this work.
This research was supported in part by the Swiss National Science Foundation
(SNF) under contracts CRSII2-141847, 200021-144352, 200020-149517,
PP00P2-128552 and by the Research Executive Agency (REA) of the European
Union under the Grant Agreements PITN--GA---2010-264564 ({\it LHCPhenoNet}),
PITN--GA--2012--316704 ({\it HiggsTools}), and the ERC Advanced Grant {\it
MC@NNLO} (340983).

\end{acknowledgments}


\begin{thebibliography}{99}


\bibitem{Aad:2012tfa}
  G.~Aad {\it et al.}  [ATLAS Collaboration],
  Phys.\ Lett.\ B {\bf 716} (2012) 1
  [arXiv:1207.7214 [hep-ex]].

\bibitem{Chatrchyan:2012ufa}
  S.~Chatrchyan {\it et al.}  [CMS Collaboration],
  Phys.\ Lett.\ B {\bf 716} (2012) 30
  [arXiv:1207.7235 [hep-ex]].


\bibitem{Abazov:2011cb}
  V.~M.~Abazov {\it et al.}  [D0 Collaboration],
  Phys.\ Rev.\ Lett.\  {\bf 108} (2012) 181803
  [arXiv:1112.0536 [hep-ex]].

\bibitem{CDFnote}
T.~Aaltonen {\it et al.} [CDF Collaboration], CDF note 11098.

\bibitem{ATLAS:2012mec}
 G.~Aad {\it et al.}  [ATLAS Collaboration],
  Phys.\ Rev.\ D {\bf 87} (2013) 11,  112001
   [Erratum-ibid.\ D {\bf 88} (2013) 7,  079906]
  [arXiv:1210.2979 [hep-ex]].

\bibitem{Chatrchyan:2013yaa}
  S.~Chatrchyan {\it et al.}  [CMS Collaboration],
  Eur.\ Phys.\ J.\ C {\bf 73} (2013) 2610
  [arXiv:1306.1126 [hep-ex]].




\bibitem{Chatrchyan:2013oev}
  S.~Chatrchyan {\it et al.}  [CMS Collaboration],
  Phys.\ Lett.\ B {\bf 721} (2013) 190
  [arXiv:1301.4698 [hep-ex]].


\bibitem{ATLAS-CONF-2014-033}
ATLAS Collaboration, ATLAS-CONF-2014-033.



\bibitem{Curtin:2014zua}
  D.~Curtin, P.~Meade and P.~J.~Tien,
  arXiv:1406.0848 [hep-ph].

\bibitem{Kim:2014eva}
  J.~S.~Kim, K.~Rolbiecki, K.~Sakurai and J.~Tattersall,
  arXiv:1406.0858 [hep-ph].

\bibitem{Luo:2014hrk}
  H.~Luo, M.~x.~Luo, K.~Wang, T.~Xu and G.~Zhu,
  arXiv:1407.4912 [hep-ph].


\bibitem{Brown:1978mq}
  R.~W.~Brown and K.~O.~Mikaelian,
  Phys.\ Rev.\ D {\bf 19} (1979) 922.


\bibitem{Ohnemus:1991kk}
  J.~Ohnemus,
  Phys.\ Rev.\ D {\bf 44} (1991) 1403.

\bibitem{Frixione:1993yp}
  S.~Frixione,
  Nucl.\ Phys.\ B {\bf 410} (1993) 280.

\bibitem{Dixon:1998py}
  L.~J.~Dixon, Z.~Kunszt and A.~Signer,
  Nucl.\ Phys.\ B {\bf 531} (1998) 3
  [hep-ph/9803250].


\bibitem{Campbell:1999ah}
  J.~M.~Campbell and R.~K.~Ellis,
  Phys.\ Rev.\ D {\bf 60} (1999) 113006
  [hep-ph/9905386].

\bibitem{Dixon:1999di}
  L.~J.~Dixon, Z.~Kunszt and A.~Signer,
  Phys.\ Rev.\ D {\bf 60} (1999) 114037
  [hep-ph/9907305].


\bibitem{Dicus:1987dj}
  D.~A.~Dicus, C.~Kao and W.~W.~Repko,
  Phys.\ Rev.\ D {\bf 36} (1987) 1570.

\bibitem{Glover:1988fe}
  E.~W.~N.~Glover and J.~J.~van der Bij,
  Phys.\ Lett.\ B {\bf 219} (1989) 488.



\bibitem{Binoth:2005ua}
  T.~Binoth, M.~Ciccolini, N.~Kauer and M.~Kramer,
  JHEP {\bf 0503} (2005) 065
  [hep-ph/0503094].

\bibitem{Binoth:2006mf}
  T.~Binoth, M.~Ciccolini, N.~Kauer and M.~Kramer,
  JHEP {\bf 0612} (2006) 046
  [hep-ph/0611170].



\bibitem{Campbell:2011cu}
  J.~M.~Campbell, R.~K.~Ellis and C.~Williams,
  JHEP {\bf 1110} (2011) 005
  [arXiv:1107.5569 [hep-ph]].


\bibitem{Campbell:2011bn}
  J.~M.~Campbell, R.~K.~Ellis and C.~Williams,
  JHEP {\bf 1107} (2011) 018
  [arXiv:1105.0020 [hep-ph]].



\bibitem{Dittmaier:2007th}
  S.~Dittmaier, S.~Kallweit and P.~Uwer,
  Phys.\ Rev.\ Lett.\  {\bf 100} (2008) 062003
  [arXiv:0710.1577 [hep-ph]].

\bibitem{Campbell:2007ev}
  J.~M.~Campbell, R.~K.~Ellis and G.~Zanderighi,
  JHEP {\bf 0712} (2007) 056
  [arXiv:0710.1832 [hep-ph]].

\bibitem{Dittmaier:2009un}
  S.~Dittmaier, S.~Kallweit and P.~Uwer,
  Nucl.\ Phys.\ B {\bf 826} (2010) 18
  [arXiv:0908.4124 [hep-ph]].



\bibitem{Melia:2011dw}
  T.~Melia, K.~Melnikov, R.~Rontsch and G.~Zanderighi,
  Phys.\ Rev.\ D {\bf 83} (2011) 114043
  [arXiv:1104.2327 [hep-ph]].


\bibitem{Greiner:2012im}
  N.~Greiner, G.~Heinrich, P.~Mastrolia, G.~Ossola, T.~\mbox{Reiter} and F.~Tramontano,
  Phys.\ Lett.\ B {\bf 713} (2012) 277
  [arXiv:1202.6004 [hep-ph]].


\bibitem{Grazzini:2005vw}
  M.~Grazzini,
  JHEP {\bf 0601} (2006) 095
  [hep-ph/0510337].

\bibitem{Wang:2013qua}
  Y.~Wang, C.~S.~Li, Z.~L.~Liu, D.~Y.~Shao and H.~T.~Li,
  Phys.\ Rev.\ D {\bf 88} (2013) 114017
  [arXiv:1307.7520].

\bibitem{Meade:2014fca}
  P.~Meade, H.~Ramani and M.~Zeng,
  arXiv:1407.4481 [hep-ph].

\bibitem{Jaiswal:2014yba}
  P.~Jaiswal and T.~Okui,
  arXiv:1407.4537 [hep-ph].


\bibitem{Dawson:2013lya}
  S.~Dawson, I.~M.~Lewis and M.~Zeng,
  Phys.\ Rev.\ D {\bf 88} (2013) 5,  054028
  [arXiv:1307.3249].





\bibitem{Bierweiler:2012kw}
  A.~Bierweiler, T.~Kasprzik, J.~H.~K{\"u}hn and S.~Uccirati,
  JHEP {\bf 1211} (2012) 093
  [arXiv:1208.3147 [hep-ph]].

\bibitem{Baglio:2013toa}
  J.~Baglio, L.~D.~Ninh and M.~M.~Weber,
  Phys.\ Rev.\ D {\bf 88} (2013) 113005
  [arXiv:1307.4331].

\bibitem{Billoni:2013aba}
  M.~Billoni, S.~Dittmaier, B.~J{\"a}ger and C.~Speckner,
  JHEP {\bf 1312} (2013) 043
  [arXiv:1310.1564 [hep-ph]].

\bibitem{Cascioli:2013gfa}
  F.~Cascioli, S.~H{\"o}che, F.~Krauss, P.~Maierh{\"o}fer, S.~Pozzorini and F.~Siegert,
  JHEP {\bf 1401} (2014) 046
  [arXiv:1309.0500 [hep-ph]].



\bibitem{Cascioli:2014yka}
  F.~Cascioli, T.~Gehrmann, M.~Grazzini, S.~Kallweit, P.~Maierh{\"o}fer, A.~von Manteuffel, S.~Pozzorini, D.~\mbox{Rathlev}, L.~Tancredi and E.\ Weihs,
 Phys.\ Lett.\ B {\bf 735} (2014) 311
  [arXiv:1405.2219 [hep-ph]].




\bibitem{Cacciari:2008gp}
  M.~Cacciari, G.~P.~Salam and G.~Soyez,
  JHEP {\bf 0804} (2008) 063
  [arXiv:0802.1189 [hep-ph]].


\bibitem{Cascioli:2013wga}
  F.~Cascioli, S.~Kallweit, P.~Maierh{\"o}fer and S.~Pozzorini,
  Eur.\ Phys.\ J.\ C {\bf 74} (2014) 2783
  [arXiv:1312.0546 [hep-ph]].

\bibitem{Denner:2012yc}
  A.~Denner, S.~Dittmaier, S.~Kallweit and S.~Pozzorini,
  JHEP {\bf 1210} (2012) 110
  [arXiv:1207.5018 [hep-ph]].





\bibitem{Cascioli:2011va}
  F.~Cascioli, P.~Maierh{\"o}fer and S.~Pozzorini,
  Phys.\ Rev.\ Lett.\  {\bf 108} (2012) 111601
  [arXiv:1111.5206 [hep-ph]].

\bibitem{collier}
A.~Denner, S.~Dittmaier and L.~Hofer, in preparation.  

\bibitem{Denner:2002ii}
  A.~Denner and S.~Dittmaier,
  Nucl.\ Phys.\ B {\bf 658} (2003) 175
  [hep-ph/0212259].


\bibitem{Denner:2005nn}
  A.~Denner and S.~Dittmaier,
  Nucl.\ Phys.\ B {\bf 734} (2006) 62
  [hep-ph/0509141].

\bibitem{Denner:2010tr}
  A.~Denner and S.~Dittmaier,
  Nucl.\ Phys.\ B {\bf 844} (2011) 199
  [arXiv:1005.2076 [hep-ph]].


\bibitem{Ossola:2006us}
  G.~Ossola, C.~G.~Papadopoulos and R.~Pittau,
  Nucl.\ Phys.\ B {\bf 763} (2007) 147
  [hep-ph/0609007].


\bibitem{Ossola:2007ax}
  G.~Ossola, C.~G.~Papadopoulos and R.~Pittau,
  JHEP {\bf 0803} (2008) 042
  [arXiv:0711.3596 [hep-ph]].



\bibitem{vanHameren:2010cp}
  A.~van Hameren,
  Comput.\ Phys.\ Commun.\  {\bf 182} (2011) 2427
  [arXiv:1007.4716 [hep-ph]].





\bibitem{Gehrmann:2013cxs}
  T.~Gehrmann, L.~Tancredi and E.~Weihs,
  JHEP {\bf 1308} (2013) 070
  [arXiv:1306.6344 [hep-ph]].


\bibitem{Henn:2014lfa}
  J.~M.~Henn, K.~Melnikov and V.~A.~Smirnov,
  JHEP {\bf 1405} (2014) 090
  [arXiv:1402.7078 [hep-ph]].


\bibitem{Gehrmann:2014bfa}
  T.~Gehrmann, A.~von Manteuffel, L.~Tancredi and E.~Weihs,
  JHEP {\bf 1406} (2014) 032
  [arXiv:1404.4853 [hep-ph]].


\bibitem{Caola:2014lpa}
  F.~Caola, J.~M.~Henn, K.~Melnikov and V.~A.~Smirnov,
  arXiv:1404.5590 [hep-ph].

\bibitem{Anastasiou:2014}
C.~Anastasiou, J.~Cancino, F.~Chavez, C.~Duhr, A.~Lazopoulos, B.~Mistlberger and R.~Mueller,
  arXiv:1408.4546 [hep-ph].

\bibitem{inprep}
T.~Gehrmann, A.~von~Manteuffel, L.~Tancredi, in preparation.

\bibitem{Chachamis:2008yb}
  G.~Chachamis, M.~Czakon and D.~Eiras,
  JHEP {\bf 0812} (2008) 003
  [arXiv:0802.4028 [hep-ph]].


\bibitem{Vollinga:2004sn}
 J.~Vollinga and S.~Weinzierl,
 Comput.\ Phys.\ Commun.\  {\bf 167} (2005) 177
 [hep-ph/0410259].

\bibitem{Bauer:2000cp}
 C.~W.~Bauer, A.~Frink and R.~Kreckel,
J.\ Symbolic\ Computation {\bf 33} (2002) 1,
 cs/0004015 [cs-sc].


\bibitem{Catani:2007vq}
  S.~Catani and M.~Grazzini,
  Phys.\ Rev.\ Lett.\  {\bf 98} (2007) 222002
[hep-ph/0703012].


\bibitem{Bozzi:2005wk}
  G.~Bozzi, S.~Catani, D.~de Florian and M.~Grazzini,
  Nucl.\ Phys.\ B {\bf 737} (2006) 73
  [hep-ph/0508068].

\bibitem{Catani:2013tia}
  S.~Catani, L.~Cieri, D.~de Florian, G.~Ferrera and M.~Grazzini,
  Nucl.\ Phys.\ B {\bf 881} (2014) 414
  [arXiv:1311.1654 [hep-ph]].



\bibitem{Catani:2011kr}
  S.~Catani and M.~Grazzini,
  Eur.\ Phys.\ J.\ C {\bf 72} (2012) 2013
   [Erratum-ibid.\ C {\bf 72} (2012) 2132]
[arXiv:1106.4652 [hep-ph]].

\bibitem{Catani:2012qa}
  S.~Catani, L.~Cieri, D.~de Florian, G.~Ferrera and M.~Grazzini,
  Eur.\ Phys.\ J.\ C {\bf 72} (2012) 2195
[arXiv:1209.0158 [hep-ph]].



\bibitem{Gehrmann:2012ze}
  T.~Gehrmann, T.~L{\"u}bbert and L.~L.~Yang,
  Phys.\ Rev.\ Lett.\  {\bf 109} (2012) 242003
[arXiv:1209.0682 [hep-ph]].

\bibitem{Gehrmann:2014yya}
  T.~Gehrmann, T.~L{\"u}bbert and L.~L.~Yang,
  JHEP {\bf 1406} (2014) 155
  [arXiv:1403.6451 [hep-ph]].



\bibitem{Catani:2009sm}
  S.~Catani, L.~Cieri, G.~Ferrera, D.~de Florian and M.~Grazzini,
  Phys.\ Rev.\ Lett.\  {\bf 103} (2009) 082001
[arXiv:0903.2120 [hep-ph]].

\bibitem{Ferrera:2011bk}
  G.~Ferrera, M.~Grazzini and F.~Tramontano,
  Phys.\ Rev.\ Lett.\  {\bf 107} (2011) 152003
[arXiv:1107.1164 [hep-ph]].

\bibitem{Catani:2011qz}
  S.~Catani, L.~Cieri, D.~de Florian, G.~Ferrera and M.~Grazzini,
  Phys.\ Rev.\ Lett.\  {\bf 108} (2012) 072001
  [arXiv:1110.2375 [hep-ph]].

\bibitem{Grazzini:2013bna}
  M.~Grazzini, S.~Kallweit, D.~Rathlev and A.~Torre,
  Phys.\ Lett.\ B {\bf 731} (2014) 204
  [arXiv:1309.7000 [hep-ph]].


\bibitem{Ferrera:2014lca}
  G.~Ferrera, M.~Grazzini and F.~Tramontano,
  arXiv:1407.4747 [hep-ph].


\bibitem{Catani:1996jh}
  S.~Catani and M.~H.~Seymour,
  Phys.\ Lett.\ B {\bf 378} (1996) 287
  [hep-ph/9602277].

\bibitem{Catani:1996vz}
S.~Catani and M.~H.~Seymour,
  Nucl.\ Phys.\ B {\bf 485} (1997) 291
   [Erratum-ibid.\ B {\bf 510} (1998) 503]
[hep-ph/9605323].


\bibitem{Martin:2010db}
  A.~D.~Martin, W.~J.~Stirling, R.~S.~Thorne and G.~Watt,
  Eur.\ Phys.\ J.\ C {\bf 70} (2010) 51
  [arXiv:1007.2624 [hep-ph]].

\bibitem{Martin:2009iq}
  A.~D.~Martin, W.~J.~Stirling, R.~S.~Thorne and G.~Watt,
  Eur.\ Phys.\ J.\ C {\bf 63} (2009) 189
[arXiv:0901.0002 [hep-ph]].



\bibitem{Heinemeyer:2013tqa}
S.~Heinemeyer {\it et al.}  [LHC Higgs Cross Section Working Group Collaboration],
arXiv:1307.1347 [hep-ph].







\end{thebibliography}
\end{document}